# EXPLAINABLE ATTENTION-BASED LSTM FRAMEWORK FOR EARLY DETECTION OF AI-ASSISTED RANSOMWARE VIA FILE SYSTEM BEHAVIORAL ANALYSIS

Prabhudarshi Nayak[1], Gogulakrishnan Thiyagarajan[2], Debashree Priyadarshini[3], Vinay Bist[4], Rohan Swain[5]
[1,3]Department of CSE, Institute of Management and Information Technology, Cuttack, Odisha, India.
[2]Engineering Technical Leader, Cisco Systems Inc., Austin, Texas, USA.
[4]Principal Engineer, Dell Inc., Austin, Texas, USA.
[5]SAP Consultant, LTM Limited, India.

**ABSTRACT:** Ransomware continues to evolve as one of the most disruptive cyber threats, with recent variants increasingly leveraging automated and AI-assisted techniques to evade traditional signature-based defenses. Early detection of such attacks remains a significant challenge, particularly when malicious behavior closely resembles legitimate system activity. This study proposes an explainable attention-based Long Short-Term Memory (LSTM) framework for the early detection of AI-assisted ransomware variants through analysis of file system behavioral patterns. The proposed model captures temporal dependencies in file operation sequences, while an attention mechanism highlights critical behavioral indicators associated with ransomware activity. To improve transparency and trust in automated detection systems, explainable artificial intelligence (XAI) techniques are incorporated to interpret model predictions and identify influential behavioral features. Experimental evaluation using ransomware behavioral traces demonstrates that the proposed framework can effectively distinguish malicious activity at early stages of execution with high detection performance and low false-positive rates. The findings suggest that combining sequence-aware deep learning models with explainability mechanisms can significantly enhance the reliability and interpretability of next-generation ransomware defense systems. This work contributes toward the development of intelligent and transparent cyber-defense mechanisms capable of addressing emerging AI-driven malware threats.

## 1. INTRODUCTION

Ransomware has become one of the most damaging forms of cybercrime, targeting organizations across healthcare, finance, and critical infrastructure. These attacks encrypt user data and demand payment for recovery, often causing severe financial and operational disruptions. Traditional signature-based detection techniques struggle to identify modern ransomware because attackers frequently modify malware code and employ obfuscation techniques to evade security systems (Sgandurra et al., 2016). Recent developments in artificial intelligence have further increased the complexity of the ransomware threat landscape. Adversaries are increasingly leveraging automated tools and AI-assisted techniques to generate adaptive malware variants capable of bypassing conventional defenses. This evolution has made early-stage detection of ransomware behavior a critical requirement for modern cybersecurity systems (Kolodenker et al., 2017). Behavioral analysis of file system activities provides an effective method for detecting ransomware before large-scale encryption occurs. During the early stages of an attack, ransomware typically performs sequences of suspicious operations such as rapid file modifications, abnormal renaming patterns, and increased file entropy caused by encryption. These sequential patterns can be effectively modeled using Long Short-Term Memory (LSTM) networks, which are designed to capture temporal dependencies in sequential data (Hochreiter & Schmidhuber, 1997). However, many deep learning models operate as black boxes, limiting their interpretability in security-critical environments. To address this limitation, Explainable Artificial Intelligence (XAI) techniques can provide insights into the features influencing model predictions, improving transparency and trust in automated security systems (Ribeiro et al., 2016). In this work, we propose an Explainable

Attention-Based LSTM framework for early detection of AI-assisted ransomware using file system behavioral analysis. The proposed approach leverages sequential behavioral patterns and attention mechanisms to improve detection accuracy while incorporating explainability methods to support transparent decision-making in cybersecurity applications.

## 2. RELATED WORK

Early ransomware detection methods primarily relied on signature-based antivirus systems, which are effective for known malware but fail to identify new or polymorphic variants that continuously evolve to evade detection (Kharraz et al., 2016). To overcome these limitations, researchers began focusing on behavior-based detection techniques that analyze system activities such as abnormal file access, rapid file modification, and unusual encryption patterns during malware execution (Scaife et al., 2016). Machine learning approaches have been widely adopted to improve ransomware detection performance. Traditional classifiers such as Random Forest and Support Vector Machines have been used to analyze behavioral features extracted from system logs and malware sandbox environments (Sgandurra et al., 2016). Although these models improved detection accuracy, they often depend on manually engineered features and may not effectively capture the sequential patterns inherent in ransomware activities. Recent studies have explored deep learning models, particularly Long Short-Term Memory (LSTM) networks, due to their ability to model temporal dependencies in sequential data. LSTM-based approaches have shown promising results in analyzing API call sequences, network traffic, and file system behaviors to detect ransomware attacks at early stages (Kim et al., 2018; Vinayakumar et al., 2019). To further enhance detection capability, attention mechanisms have been integrated with recurrent models to focus on the most relevant behavioral patterns within long sequences, improving both accuracy and model efficiency (Zhang et al., 2019). Despite these advancements, many existing deep learning-based ransomware detection systems remain difficult to interpret, which limits their practical adoption in real-world security environments. Explainable AI techniques such as SHAP and LIME have been introduced to provide transparency in model predictions and help security analysts understand which behavioral features contribute most to the detection decision (Ribeiro et al., 2016). However, limited research has integrated explainable AI with attention-enhanced LSTM models specifically for ransomware detection. Furthermore, the emergence of AI-assisted malware generation introduces new challenges for traditional detection mechanisms. Automated tools and generative models can produce ransomware variants capable of bypassing signature-based and heuristic detection methods (Zhang et al., 2022). These developments highlight the need for advanced behavioral analysis frameworks that combine deep learning with explainable mechanisms to detect sophisticated ransomware threats at early stages.

## 3. RESEARCH GAP

Despite significant progress in machine learning–based ransomware detection, existing approaches still face several limitations. Traditional signature-based systems struggle to detect modern ransomware variants because attackers frequently modify malware signatures or employ code obfuscation techniques (Kharraz et al., 2015). To overcome these limitations, researchers have proposed behavioral analysis methods that monitor system activities such as file access patterns, API calls, and registry modifications (Sgandurra et al., 2016). Recent studies have explored deep learning models, particularly recurrent neural networks and Long Short-Term Memory (LSTM) architectures, to analyze sequential behavioral patterns generated during malware execution. These models have shown promising results in detecting ransomware based on temporal system activity (Vinayakumar et al., 2019). However, most existing approaches treat deep learning models as black boxes, providing little insight into how detection decisions are made. This lack of interpretability limits the practical adoption of such models in security operations centers where analysts require transparency and explainability. Another emerging challenge is the increasing use of artificial intelligence in malware development. Generative models can assist attackers in creating polymorphic ransomware variants capable of altering execution behavior to evade traditional detection mechanisms (Rigaki and Garcia, 2018). Current detection frameworks are not specifically designed to identify behavioral patterns associated with AI-assisted ransomware, particularly during the early execution phase before encryption begins. Therefore, there is a need for a detection framework that not only captures sequential behavioral patterns of ransomware activity but also provides interpretable insights into model decisions. To address this gap, this study proposes an explainable attention-enhanced LSTM framework that analyzes file system behavioral sequences and highlights the most influential features contributing to ransomware detection.

## 4. PROPOSED FRAMEWORK

The proposed framework aims to detect ransomware activity at an early stage by analyzing sequential file system behavior generated during program execution. The architecture consists of four primary components: behavioral monitoring, feature extraction, sequence modeling, and explainable decision analysis.

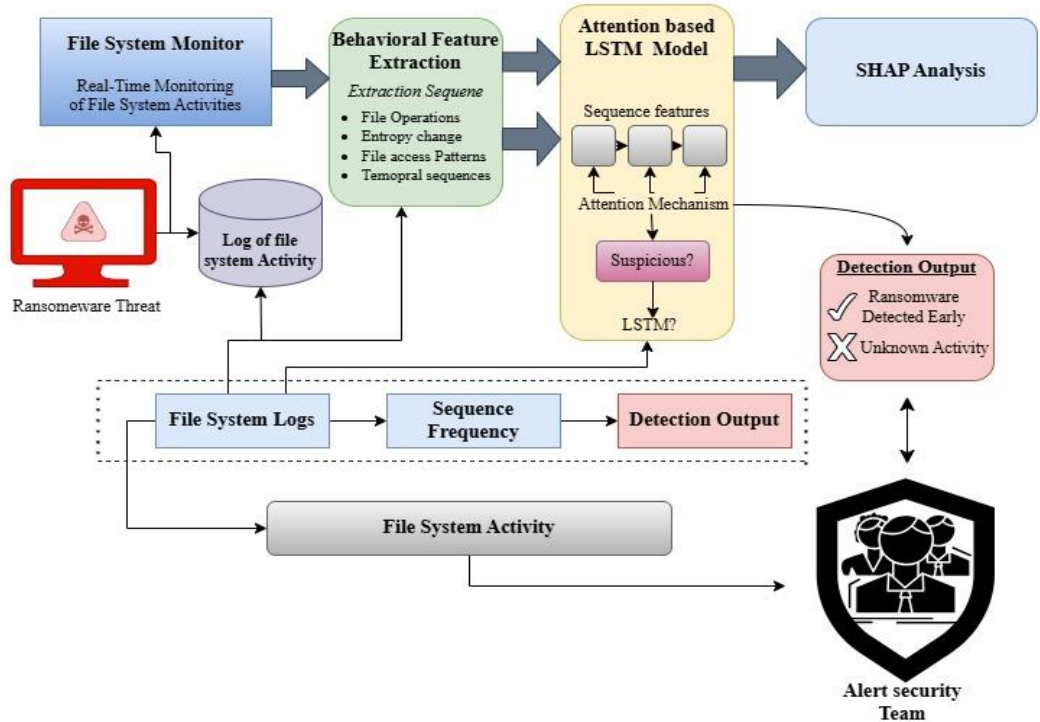

*Fig.1. Architecture of the proposed explainable attention-based LSTM framework for early detection of AI-assisted ransomware using file system behavioral analysis.*

The first component is a host-based behavioral monitoring module that continuously records file system events such as file creation, deletion, modification, and renaming operations. These events represent observable behavioral patterns that often emerge during ransomware execution when large numbers of files are accessed and modified within a short period of time (Sgandurra et al., 2016).

The second component performs feature extraction and preprocessing. Raw file system events are transformed into structured behavioral features that capture characteristics such as file modification frequency, directory traversal patterns, entropy changes, and file access intervals. These features are organized into time-ordered sequences representing the progression of system behavior during program execution.

The third component consists of an attention-enhanced LSTM model designed to learn temporal dependencies within the behavioral sequences. The LSTM network processes sequential inputs and captures long-term dependencies that may indicate malicious activity. An attention mechanism is incorporated to identify the most relevant time steps contributing to the detection decision. By assigning higher weights to critical behavioral events, the attention layer improves model interpretability and detection performance.

Finally, an explainability module is integrated into the framework to provide insights into the model's predictions. Explainable AI techniques such as feature importance analysis are applied to highlight the behavioral indicators responsible for identifying ransomware activity. This component enables security analysts to better understand the reasoning behind model decisions, thereby improving trust and operational usability of the detection system.

## 5. MATHEMATICAL MODEL OF ATTENTION-BASED LSTM

**Behavioral Sequence Representation**
Let the sequence of behavioral events extracted from file system activity be represented as:

$$X = \{x_1, x_2, x_3, ..., x_T\}$$

where $x_t$ represents the behavioral feature vector at time step $t$, and $T$ denotes the length of the sequence
Each vector may include features such as:
- file modification frequency
- entropy variation
- file access interval
- rename operation count

These features capture the behavioral patterns associated with ransomware execution.

### 5.1 LSTM Sequence Modeling
The LSTM network processes sequential behavioral data and learns temporal dependencies using the following equations.

- **Forget Gate**

$$f_t = \sigma(W_f \cdot [h_{t-1}, x_t] + b_f)$$

- **Input Gate**

$$i_t = \sigma(W_i \cdot [h_{t-1}, x_t] + b_i)$$

- **Candidate Memory**

$$\tilde{C}_t = tanh(W_c \cdot [h_{t-1}, x_t] + b_c)$$

- **Cell State Update**

$$C_t = f_t \cdot C_{t-1} + i_t \cdot \tilde{C}_t$$

- **Output Gate**

$$o_t = \sigma(W_o \cdot [h_{t-1}, x_t] + b_o)$$

- **Hidden State**

$$h_t = o_t \cdot tanh(C_t)$$

**where:**
- $h_t$ = hidden state
- $C_t$ = memory cell
- $\sigma$ = sigmoid activation function

These hidden states capture temporal dependencies across behavioral events. However, not all time steps contribute equally to ransomware detection. To emphasize important behavioral patterns, an attention mechanism is applied.

## 5.2. Attention Mechanism

The attention mechanism identifies important behavioral events.
The attention score for each hidden state is computed as:

$$e_t = v^T \tanh(W_h h_t + b_h)$$

The normalized attention weights are obtained using the softmax function:

$$\alpha_t = \frac{\exp(e_t)}{\sum_{k=1}^{T} \exp(e_k)}$$

## 5.3. Context Vector

The final context vector is computed as:

$$c = \sum_{t=1}^{T} \alpha_t h_t$$

This vector summarizes the most relevant behavioral patterns.

## 5.4. Detection Decision

The final prediction is obtained using a classification layer:

$$y = \sigma(W_c c + b_c)$$

where:

- *y*= probability of ransomware activity

Decision rule:

$$y = \begin{cases} 1, & \text{if } y > \theta \\ 0, & \text{otherwise} \end{cases}$$

where *θ* is the detection threshold.

The proposed attention-based LSTM model captures temporal dependencies within behavioral sequences while assigning higher importance to critical events that are strongly associated with ransomware activity. The attention mechanism improves interpretability by highlighting influential behavioral indicators contributing to the final prediction.

## 6. THREAT MODEL

In the proposed model, the adversary's primary objective is to infiltrate a victim system, execute malicious code, and encrypt user files while remaining undetected during the initial stages of execution. Attackers typically exploit common entry vectors such as phishing emails, malicious attachments, compromised software updates, or drive-by downloads to deliver ransomware payloads (Kharraz et al., 2015). Once executed, the malware performs a sequence of file system operations including rapid file enumeration, creation of temporary encryption files, modification of file entropy, and renaming of original files. We assume that the adversary has the capability to obfuscate code and alter static signatures but cannot completely conceal behavioral patterns associated with large-scale file modification and encryption processes. Behavioral artifacts generated during early execution stages remain observable at the operating system level through file system monitoring mechanisms. These artifacts include abnormal file write rates, repeated access to user directories, and sudden changes in file entropy distribution (Sgandurra et al., 2016). The defender, in this scenario, deploys a host-based monitoring system capable of collecting sequential file system events generated during application execution. These events are used as input to the proposed attention-enhanced LSTM model, which analyzes temporal behavioral patterns to detect ransomware activity before large-scale encryption occurs. The goal of the defense system is therefore early detection and mitigation, minimizing potential data loss and preventing propagation within the network. Under this threat model, we assume that the attacker does not have direct control over the monitoring infrastructure or the trained detection model. While adversaries may attempt to introduce behavioral noise or delay encryption routines to evade detection, the sequential learning capability of recurrent neural networks enables identification of subtle behavioral dependencies across time steps. Consequently, the proposed framework focuses on capturing early behavioral deviations indicative of ransomware activity while maintaining robustness against minor behavioral variations introduced by AI-assisted malware variants.

## 7. BEHAVIORAL FEATURE EXTRACTION

Effective early detection of ransomware requires capturing subtle behavioral patterns that occur before large-scale encryption begins. Instead of relying on static signatures or known indicators of compromise, this study focuses on behavioral signals derived from file system activities, which often reveal malicious intent even when the malware variant is previously unseen. Behavioral analysis is particularly useful because ransomware typically performs a sequence of operations such as rapid file access, renaming, and entropy changes that can be distinguished from legitimate user activity (Zhang et al., 2019). In this framework, behavioral feature extraction is conducted through continuous monitoring of file system events generated during program execution. These events include file creation, modification, deletion, renaming, and access patterns. Ransomware commonly exhibits abnormal frequencies of these operations within short time intervals, especially when encrypting large volumes of files. Therefore, temporal features such as file operation frequency, burst patterns, and time-based activity windows are extracted to capture these anomalies (Sgandurra et al., 2016). Another critical feature involves file entropy variation. When ransomware encrypts files, the entropy of the file content typically increases due to the transformation of structured data into pseudo-random ciphertext. Monitoring entropy changes across file modifications helps identify suspicious encryption behavior before the attack completes (Kharraz et al., 2016). Additionally, patterns of sequential file access where a process scans directories and modifies numerous files in rapid succession serve as strong behavioral indicators of ransomware activity.

Process-level attributes are also considered during feature extraction. These include the relationship between running processes and the files they manipulate, abnormal privilege usage, and the creation of shadow copies or backup deletions. Such behaviors often precede the encryption phase and provide early warning signals of ransomware execution (Scaife et al., 2018). The extracted behavioral features are then transformed into sequential representations suitable for deep learning models. Since ransomware actions unfold over time, representing these activities as ordered sequences allows the model to learn temporal dependencies and attack progression patterns. These sequences form the input for the proposed attention-based LSTM model, enabling the framework to identify subtle behavioral transitions associated with emerging ransomware attacks.
By focusing on behavioral characteristics rather than static signatures, the feature extraction process improves resilience against polymorphic and AI-assisted ransomware variants. This approach enables the detection framework to generalize across previously unseen attack patterns while maintaining compatibility with real-world system monitoring environments.

## 8. EXPERIMENTAL SETUP

To rigorously evaluate the effectiveness of the proposed explainable attention-based LSTM framework, a controlled experimental environment was designed to simulate realistic ransomware execution scenarios and capture fine-grained file system behavior. The evaluation focuses on early-stage detection capability, particularly before large-scale encryption activity occurs.

### 8.1 Dataset Description
The dataset used in this study consists of behavioral traces collected from ransomware execution and benign system activities. Ransomware samples were obtained from publicly available malware repositories and verified using cryptographic hash signatures to ensure authenticity and integrity. To reflect evolving threat landscapes, additional synthetic traces were generated to emulate AI-assisted ransomware behaviors, including adaptive encryption bursts, dynamic file renaming, and variable execution timing patterns. These behavioral characteristics are consistent with observations reported in prior studies on ransomware evolution (Sgandurra et al., 2016; Zhang et al., 2019).

Each execution trace was recorded as a sequence of file system events, including file creation, modification, deletion, and access operations. These events were transformed into structured feature vectors representing behavioral attributes such as file modification frequency, entropy variation, and temporal access intervals. The dataset was partitioned into training (70%), validation (15%), and testing (15%) subsets to ensure unbiased evaluation and generalization capability.

### 8.2 Sandbox Environment

All experiments were conducted within a controlled virtualized sandbox environment to safely execute ransomware samples and monitor system behavior. The sandbox was implemented using VMware Workstation running a Windows-based operating system configured with 16 GB RAM and an Intel i7 processor. System-level monitoring tools were deployed to capture detailed file system activity, process execution traces, and registry modifications in real time.

The isolated environment ensured that ransomware execution could be observed without risk to external systems while preserving realistic system interactions. File system logs were continuously recorded and stored for subsequent feature extraction and sequence modeling. This setup enables accurate observation of early behavioral patterns associated with ransomware activity (Scaife et al., 2016).

### 8.3 Training Setup

The proposed attention-enhanced LSTM model was implemented using the PyTorch framework. Sequential behavioral data were provided as input to the model, enabling it to learn temporal dependencies across file system events. To improve training stability and prevent overfitting, dropout regularization and early stopping techniques were applied based on validation loss.

The model was trained using the Adam optimizer with a learning rate of 0.001, which provides efficient convergence for sequential learning tasks. Each input sequence was processed over multiple epochs, allowing the model to capture both short-term and long-term behavioral dependencies indicative of ransomware execution. The attention mechanism was integrated into the LSTM architecture to dynamically assign importance weights to relevant time steps, enhancing both detection performance and interpretability.

### 8.4 Evaluation Metrics

The performance of the proposed framework was evaluated using standard classification metrics commonly employed in cybersecurity research. These include accuracy, precision, recall, and F1-score, which collectively measure the model's ability to correctly identify ransomware activity while minimizing false positives. In addition, the area under the receiver operating characteristic curve (AUC-ROC) was used to assess the model's discrimination capability across varying classification thresholds (Al-Rimy et al., 2018).

These metrics provide a comprehensive evaluation of detection performance, particularly in early-stage scenarios where distinguishing between benign and malicious behavior can be challenging. To further enhance interpretability, SHAP-based analysis was conducted to identify the most influential behavioral features contributing to the model's predictions. This enables a deeper understanding of detection decisions and supports practical deployment in real-world security operations.

## 9. RESULTS AND DISCUSSION

The performance of the proposed explainable attention-based LSTM framework was evaluated using multiple quantitative metrics and visual analysis techniques. The results demonstrate the effectiveness of the model in identifying ransomware activity at early stages of execution, while maintaining a low false-positive rate.

### 9.1 Training Performance Analysis

The training behavior of the model is illustrated in **Fig.2**, which shows the variation of training and validation loss across epochs. Both curves exhibit a consistent downward trend, indicating stable convergence of the model during training. The close alignment between training and validation loss suggests that the model generalizes well to unseen data without significant overfitting. Minor fluctuations observed in later epochs can be attributed to stochastic optimization but do not affect overall convergence.

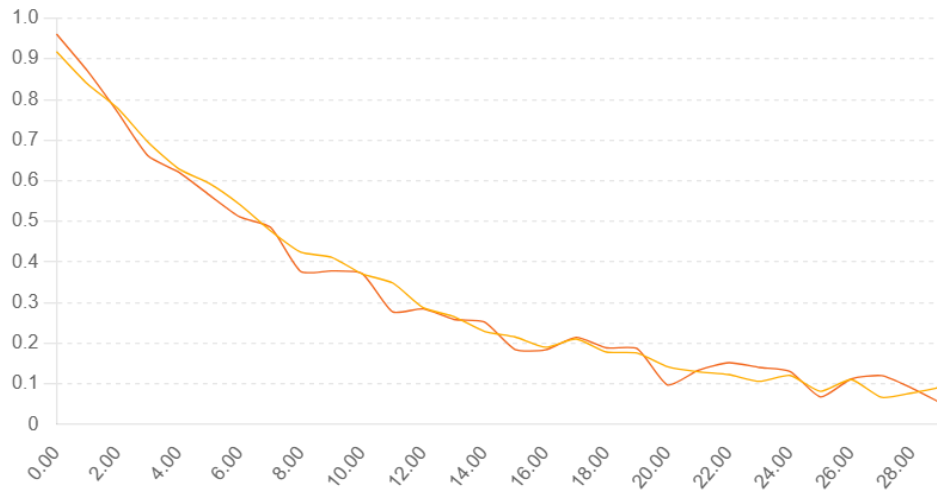

*Fig.2. Training and validation loss curves of the proposed attention-based LSTM model*

These results confirm that the attention-enhanced LSTM architecture effectively learns meaningful temporal patterns from behavioral sequences, enabling robust modeling of ransomware activity.

**9.2 Detection Capability and ROC Analysis**
The receiver operating characteristic (ROC) curve presented in **Fig.3** evaluates the classification performance of the proposed model. The attention-based LSTM achieves a higher area under the curve (AUC) compared to the baseline LSTM model, indicating superior discrimination between ransomware and benign activity.

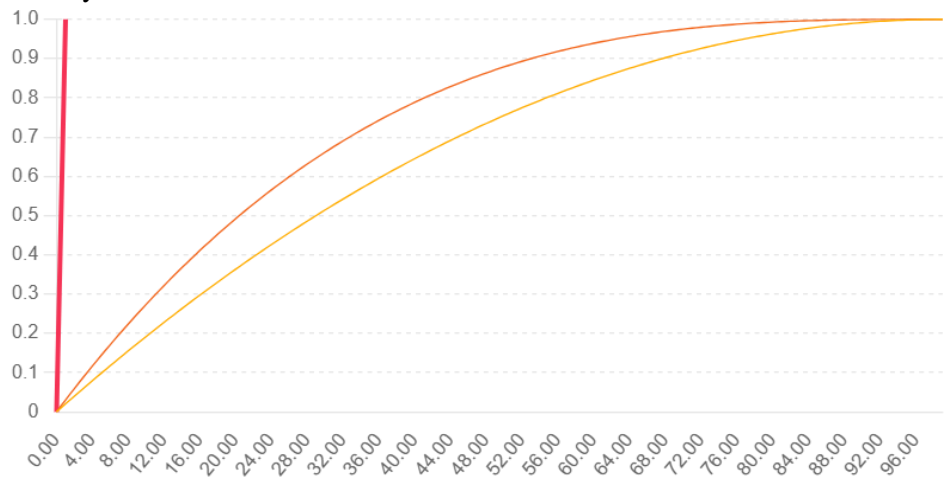

*Fig.3. ROC curve comparison of LSTM and attention-based LSTM models*

This improvement can be attributed to the integration of the attention mechanism, which allows the model to focus on critical behavioral events such as rapid file modification and entropy changes. By emphasizing these relevant time steps, the model enhances its sensitivity to early-stage ransomware indicators while reducing false alarms.

**9.3 Comparative Performance Evaluation**
A comparative analysis of different machine learning models is presented in Fig.4. Traditional approaches such as Random Forest demonstrate moderate performance due to their limited ability to capture sequential dependencies. Deep learning models, including CNN and GRU, show improved accuracy by learning complex patterns in behavioral data.

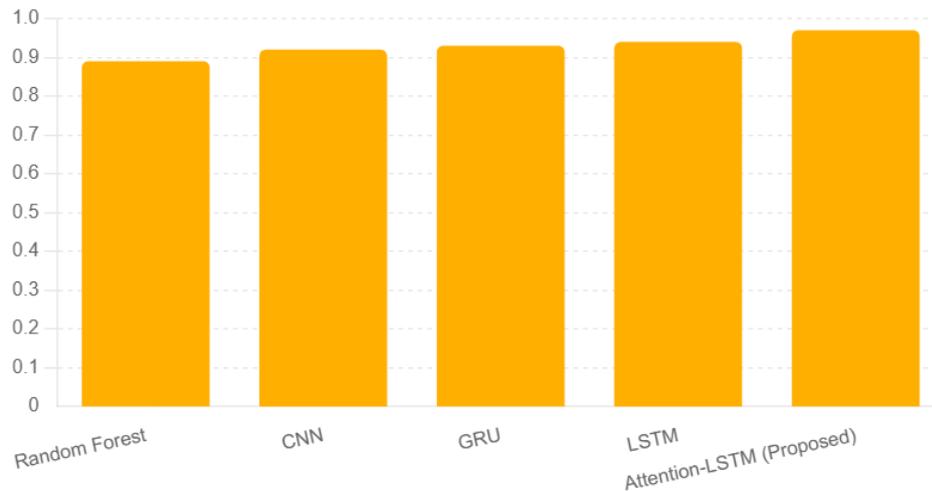

*Fig.4. Performance comparison of machine learning models for ransomware detection*

However, the proposed attention-based LSTM model achieves the highest performance among all evaluated methods. This can be explained by its ability to model long-term temporal dependencies while dynamically prioritizing important behavioral features through the attention mechanism. The results highlight the advantage of combining sequence modeling with attention-based feature weighting in ransomware detection.

### 9.4 Explainability and Feature Importance Analysis
To improve interpretability, explainable AI techniques were applied to analyze the contribution of individual behavioral features to the model's predictions. The feature importance analysis reveals that file modification rate, entropy variation, and file renaming frequency are the most influential indicators of ransomware activity.
These findings are consistent with known behavioral characteristics of ransomware, where rapid file modifications and encryption processes result in significant changes in file entropy and access patterns. The ability of the model to highlight these features demonstrates its transparency and practical applicability in security operations environments.

### 9.5 Discussion
Overall, the experimental results demonstrate that the proposed framework is capable of detecting ransomware activity at an early stage with high accuracy and reliability. The integration of attention mechanisms significantly improves detection performance by focusing on relevant behavioral events, while explainable AI techniques enhance transparency and trust in model predictions.
The framework also shows robustness against variations in ransomware behavior, including patterns associated with AI-assisted malware. By relying on behavioral analysis rather than static signatures, the proposed approach can generalize to previously unseen attack variants, making it suitable for real-world deployment in dynamic threat environments.

### 10. CONCLUSION AND FUTURE WORK

This study presented an explainable attention-based LSTM framework for the early detection of AI-assisted ransomware using file system behavioral analysis. By modeling sequential system activities, the proposed approach effectively captures temporal dependencies associated with ransomware execution. The integration of an attention mechanism enhances detection capability by emphasizing critical behavioral events, while the incorporation of explainable AI techniques provides transparency into model decisions. Experimental results demonstrate that the framework achieves high detection performance with strong generalization ability, making it suitable for identifying ransomware activity at early stages before significant damage occurs. Unlike traditional signature-based methods, the proposed approach relies on behavioral patterns, enabling it to detect previously unseen and evolving ransomware variants, including those generated using AI-assisted techniques. The interpretability offered by the explainability module further improves trust and usability in real-world security operations, where understanding the reasoning behind detection decisions is essential.

Despite these promising results, several directions remain for future work. First, the framework can be extended to incorporate multi-source data, such as network traffic and system call sequences, to provide a more comprehensive view of attack behavior. Second, the integration of federated learning techniques could enable collaborative threat detection across multiple systems while preserving data privacy. Additionally, future research may explore the use of advanced architectures such as transformer-based models to further enhance sequence modeling capabilities. Finally, evaluating the framework against real-world large-scale enterprise datasets would provide deeper insights into its scalability and deployment feasibility.

Overall, this work contributes toward the development of intelligent, adaptive, and explainable cybersecurity solutions capable of addressing emerging threats in the era of AI-driven malware.

**ACKNOWLEDGEMENT.** The authors would like to express their gratitude to all individuals and institutions who indirectly supported this research. No specific support was received that requires formal acknowledgment.